\documentclass[]{spie}  

\usepackage{amsmath,amsfonts,amssymb}
\usepackage{graphicx}
\usepackage[allcolors=blue]{hyperref}

\title{VSTPOL: making the VST a large survey telescope for optical polarimetry}

\author[a]{P. Schipani}
\author[b]{S. Covino}
\author[c]{F. Snik}
\author[a]{M. Colapietro}
\author[a]{F. Perrotta}
\author[a]{S. Savarese}
\author[d]{S. Bagnulo}
\author[b]{P. Bellutti}
\author[a]{G. Capasso}
\author[f]{E. Cappellaro}
\author[g]{M. Cappi}
\author[g]{G. Castignani}
\author[a]{S. D'Orsi}
\author[f]{J. Farinato}
\author[h]{O. Hainaut}
\author[k]{D. Hutsemekers}
\author[c]{K. Kuijken}
\author[l]{A. M. Magalhaes}
\author[f]{D. Magrin}
\author[a]{M. Marconi}
\author[a]{L. Marty}
\author[h]{F. Patat}
\author[g]{E. Pian}
\author[b]{F. Rigamonti}
\author[a]{V. Ripepi}
\author[m]{P. Rossettini}
\author[a]{R. Z. Sanchez}
\author[n]{A. Smette}
\author[m]{P. Spanò}
\author[m]{R. Tomelleri}
\author[q]{G. Umbriaco}
\author[c]{A. van Vorstenbosch}
\author[r]{G. Verdoes-Kleijn}

\affil[a]{INAF - Osservatorio Astronomico di Capodimonte, Salita Moiariello 16, I-80131, Naples, Italy }
\affil[b]{INAF - Osservatorio Astronomico di Brera, Via Bianchi 46, I-23807, Merate, Italy }
\affil[c]{Leiden Observatory, Leiden University, P.O. Box 9513, 2300 RA Leiden, The Netherlands}
\affil[d]{Armagh Observatory, College Hill, Armagh, BT61 9DG, Northern Ireland, UK}
\affil[f]{INAF - Osservatorio Astronomico di Padova, Vicolo dell’Osservatorio 5, I-35122, Padua, Italy }
\affil[g]{INAF - Osservatorio di Astrofisica e Scienza dello Spazio di Bologna, Via Gobetti 93/3, I-40129 Bologna, Italy}
\affil[h]{ESO, Karl-Schwarzschild-Strasse 2, Garching bei München, D-85748, Germany}
\affil[k]{Université de Liège, Allée du 6 Août 19c, B5c, 4000, Liège, Belgium}
\affil[l]{Universidade de São Paulo, R. do Matão, 1226, São Paulo, SP 05508-090, Brazil}
\affil[m]{Tomelleri s.r.l., Viale del Lavoro 12/a, I-37069, Villafranca (VR), Italy}
\affil[n]{ESO, Alonso de Cordova 3107, Vitacura, Santiago, Chile}
\affil[q]{Università di Bologna Alma Mater Studiorum, via Piero Gobetti 93/2, I-40129,
Bologna, Italy}
\affil[r]{Kapteyn Astronomical Institute, University of Groningen, PO Box 800, 9700 AV Groningen, The Netherlands}

\authorinfo{Send correspondence to: pietro.schipani@inaf.it}

\begin{document} 
\maketitle

\begin{abstract}
Since the start of operations in 2011, the  VLT Survey Telescope (VST) has been one of the most efficient wide-field imagers in the optical bands. However, in the next years the Vera C. Rubin Observatory Legacy Survey of Space and Time (LSST) will be a game-changer in this field. Hence, the timing is appropriate for specializing the VST with additions that can make it unique in well-defined scientific cases. VSTPOL is a project that aims to provide the addition of wide-field polarimetric capabilities to the VST telescope, making it the first large survey telescope for linear optical polarimetry. Actually, while there are quite a number of optical telescopes, the telescopes providing polarimetric instrumentation are just a few. The number of relatively large mirror polarimetric telescopes is small, although they would be specifically needed e.g.~to support many science cases of the Cherenkov Telescope Array (CTA) that, in the southern hemisphere, is co-located with the VST.  
The VST telescope is equipped with a single instrument, the OmegaCAM wide-field imaging camera operating in the visible bands with a field of view of 1$^\circ\times1^\circ$. The polarimetric mode will be implemented through the insertion of a large rotatable polarizer installed on the field-corrector optics, which will be exchangeable with the non-polarimetric corrector optics. The limiting polarimetric systematic errors due to variable atmospheric conditions and instrumental polarization can be corrected down to a level of $\sim10^{-3}$ by leveraging the large amount of unpolarized stars within each field-of-view. By the user point of view, VSTPOL will be an additional mode for the VST wide-field imaging camera.
\end{abstract}


\section{INTRODUCTION}
\label{sec:intro}  

Polarization provides vital insights into astrophysical processes. Any process that breaks a symmetry in a radiative source or between an observer and a source produces polarization. For example, scattering processes cause linear polarization. This includes light scattering from dust, electron scattering and surface scattering. But also magnetic fields uniquely affect the polarization of light. Conversely, polarimetry is the best method to characterize scattering media and to measure the direction and strength of magnetic fields anywhere in the universe in the optical domain. Both magnetic fields and scattering processes occur everywhere in astronomy, and form the main theme of polarimetric science cases.

Making a large polarimetric survey telescope is an excellent opportunity to explore polarimetric survey science cases which are not achievable with the current polarimetric instruments, characterized by a small field of view (FoV). The large FoV of the VST (1$^\circ\times1^\circ$) allows for:

\begin{itemize}
    \item the observations of extended sources with a minimal amount of measurements;
    \item the simultaneous observations of multiple sources;
    \item fast follow-ups of CTA transients of which no accurate sky-coordinates are known. By being able to target a large portion of the sky surrounding the estimated origin of the signal, the chance of detection is greatly improved;
    \item surveys of notable fields, equatorial or in the southern hemisphere (e.g. COSMOS, {\it Euclid} deep fields, LSST drilling fields);
    \item serendipitous discoveries.
\end{itemize}

In addition to the large FoV, OmegaCAM also has a high spatial resolution at 0.21” that, together with the excellent Paranal observing conditions and telescope performance, guarantees high-quality imaging. 

We now quickly review some of the main science topics addressable by a facility as the VSTPOL. We have separated for simplicity two main subjects: science in connections with the CTA and science for the wide-field case. However, in several cases, the separation is rather arbitrary and is only useful for a rough classification.

\begin{figure} [ht]
\begin{center}
\begin{tabular}{c} 
\includegraphics[height=12cm]{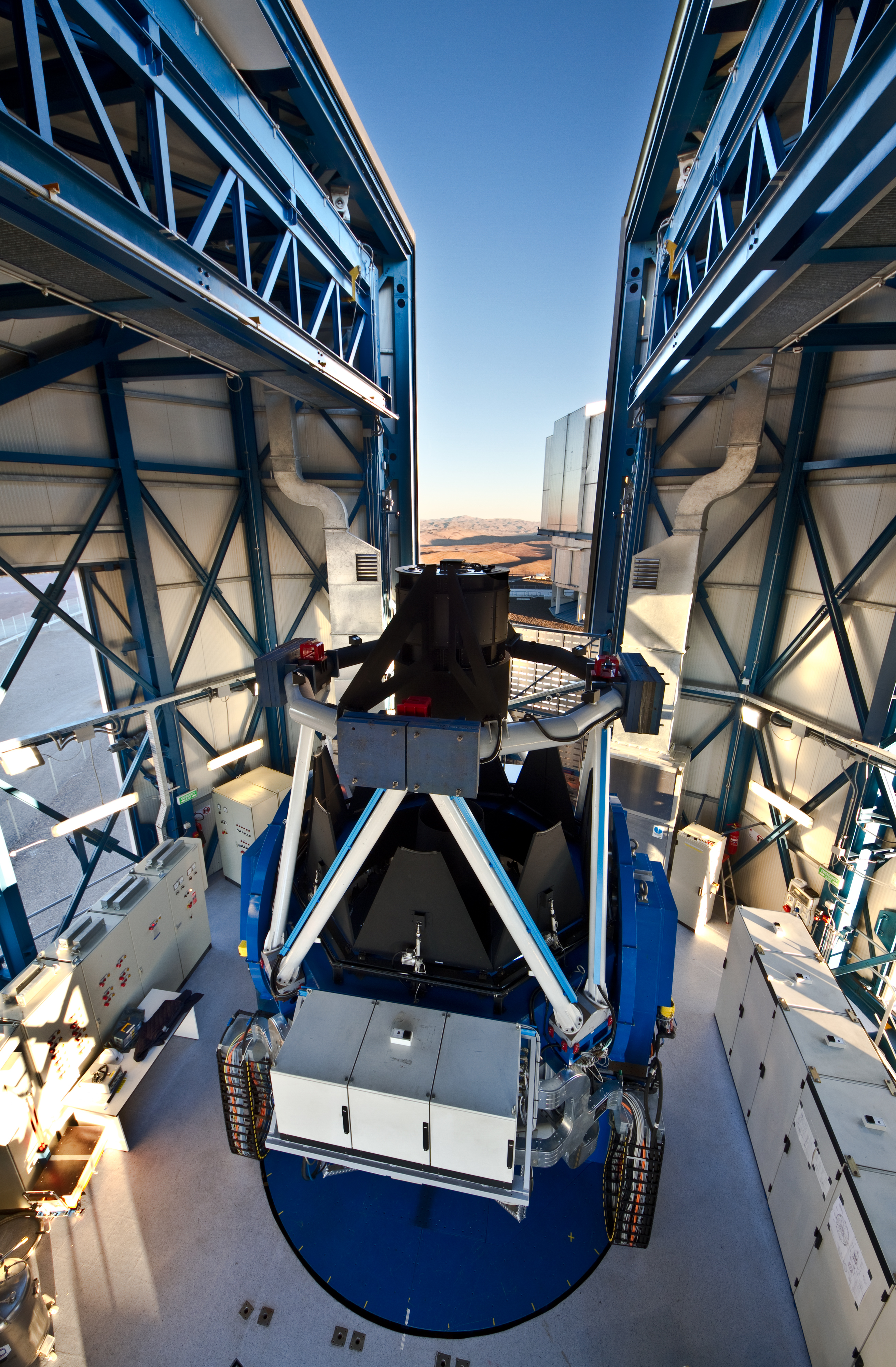}
\includegraphics[height=12cm]{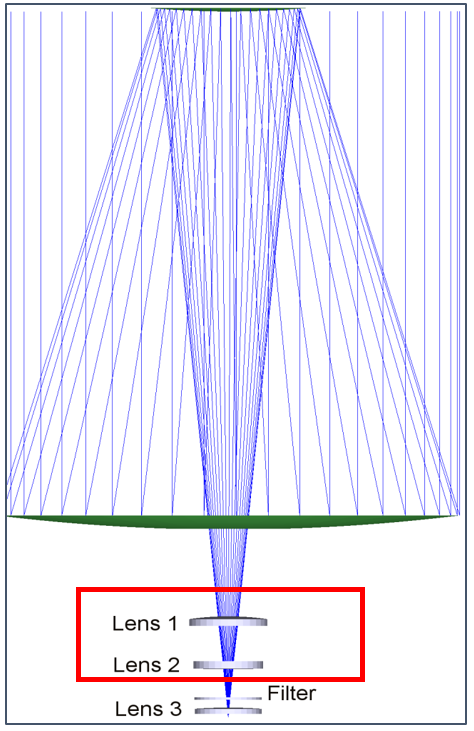}
\end{tabular}
\end{center}
\caption[example1] 
{ \label{fig:Telescope} 
Left: The VST in the Paranal Observatory. Right: optical configuration, 2-mirror design with a field-corrector composed by 2 lenses in the telescope and the camera window.}
\end{figure}

\begin{figure} [ht]
\begin{center}
\begin{tabular}{c} 
\includegraphics[height=7.5cm]{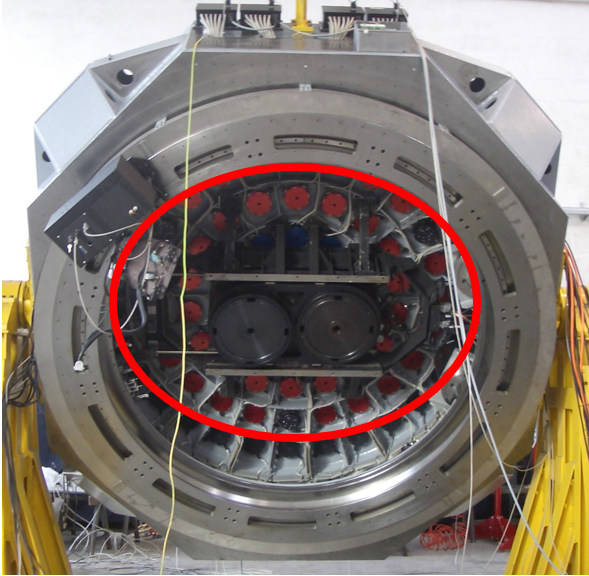}
\includegraphics[height=7.5cm]{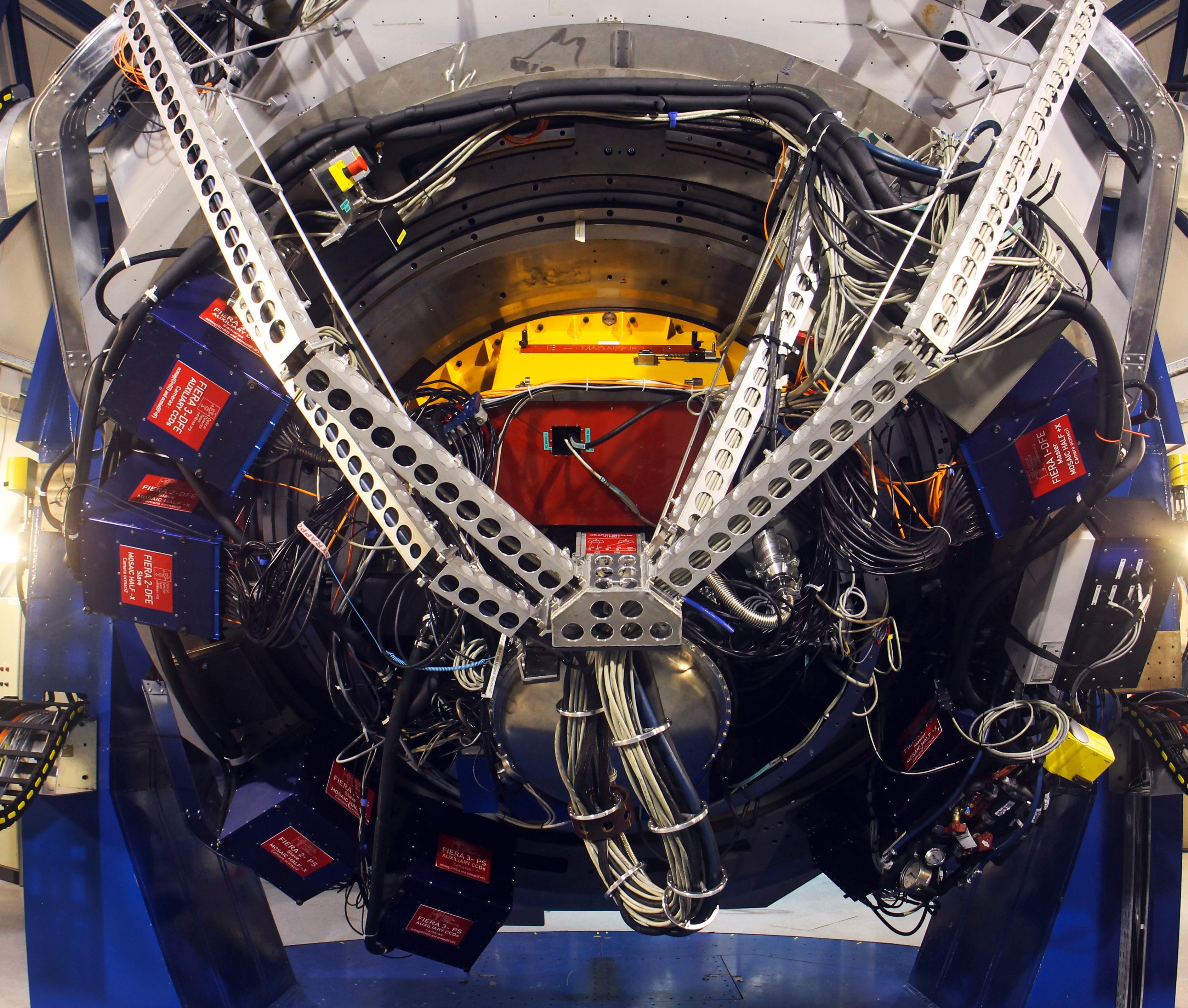}
\end{tabular}
\end{center}
\caption[example1] 
{ \label{fig:CurrentMechanics} 
Left: inner view of the VST telescope after removal of OmegaCAM instrument and telescope probe. Right: external view of the telescope in the same position with OmegaCAM installed.}
\end{figure}

\section{Synergy with CTA}
The location on the Cerro Paranal observatory site allows for some of the best seeing and light pollution conditions in the world. This makes the VST site ideal for the polarimetric “photon hungry” technique. The site is also ideal for its colocation with CTA South. In fact, the discovery space for CTA might be significantly negatively impacted without companion observations providing optical photometry and polarimetry, so options for optical telescopes capable of supporting CTA science have been explored, coming to the conclusion that, especially in the Southern hemisphere, there is a lack of viable options. Science cases where CTA mostly needs an optical and polarimetric support are e.g.: 

\begin{itemize}
    \item Active Galactic Nuclei (AGN), including blazars
    \item Transients
    \item Gamma Ray Bursts (GRBs)
    \item Any source with efficient particle acceleration
\end{itemize}

\paragraph{AGN, Blazars and generic VHE sources.} As a general rule, and not only for AGN and subclasses, at very high-energy (VHE) the emission is non-thermal and is often due to particle acceleration. Independently of the specific scenario and source physics, acceleration is usually driven by magnetic fields that introduce clear asymmetries in the sources that can only be studied by means of polarimetry. As a matter of fact, polarimetry is always a highly needed ancillary dataset for most, if not all, observing campaigns with present day and future Cherenkov telescopes. However, at VHE, many targets of interests are bright enough to show variability on short time-scales, even down to a few minutes. A polarimetric monitoring is occasionally secured by small-size facilities, able to provide a measurement per day or per week. This is unfortunately far from being optimal, and a larger-scale facility as the VSTPOL will be able to provide in most cases almost real-time polarimetric monitoring offering unprecedented coverage and modeling perspectives. 

The large field of view is clearly unnecessary for single target studies when the position of the target is accurately known, yet it can play a role studying alignments of AGN axes with their host large-scale structures. AGN linear polarization is related to the AGN/SMBH (SuperMassive Black Hole) symmetry axis and constitutes a key quantity to probe their inner structure, otherwise spatially unresolved \cite{Hutsemekeretal2014} as well as physical processes that could modify polarization over cosmological distances.

\paragraph{Transients and GRBs.} Because of the co-location of VST and CTA South, transients can be observed from the same site, under the same observing conditions, and virtually at the same time.

As for the CTA support, the wide-field of the VST is an evident advantage. In fact, the most interesting transients are rare events; when they happen, it is essential to observe their evolution since the start.  With a small FoV the chance of simultaneously observing the start of a transient with both the CTA and an optical support telescope is smaller, and this would severely impact the scientific value of the optical support. In all cases, where the transient coordinates are still not precisely known, the importance of a wide FoV is paramount. 

In a more general scenario, and this holds particularly for high-energy transients and GRBs \cite{Covino&Gotz2016}, the alert for a new transient is likely coming from a high-energy satellite. It is difficult to generalise, yet a large fraction of these events will be localized only with a relatively modest accuracy. A large-field-of-view polarimeter, with an adequate collecting area, offers the unprecedented and exciting possibility to begin acquiring data in a field when the optical transient (if any, of course) is still to be identified. This is true in particular for GRBs, and we expect 1-2 new events promptly observable from Paranal per month. But the inventory of possible transients with a rapid early time evolution is large, including the frankly unlikely but enormously potentially important case of kilonovae followed-up during the first few hours after a Gravitational Wave (GW) event \cite{Covinoetal2017}. 

\begin{figure} [ht]
\begin{center}
\begin{tabular}{c} 
\includegraphics[height=7.75cm]{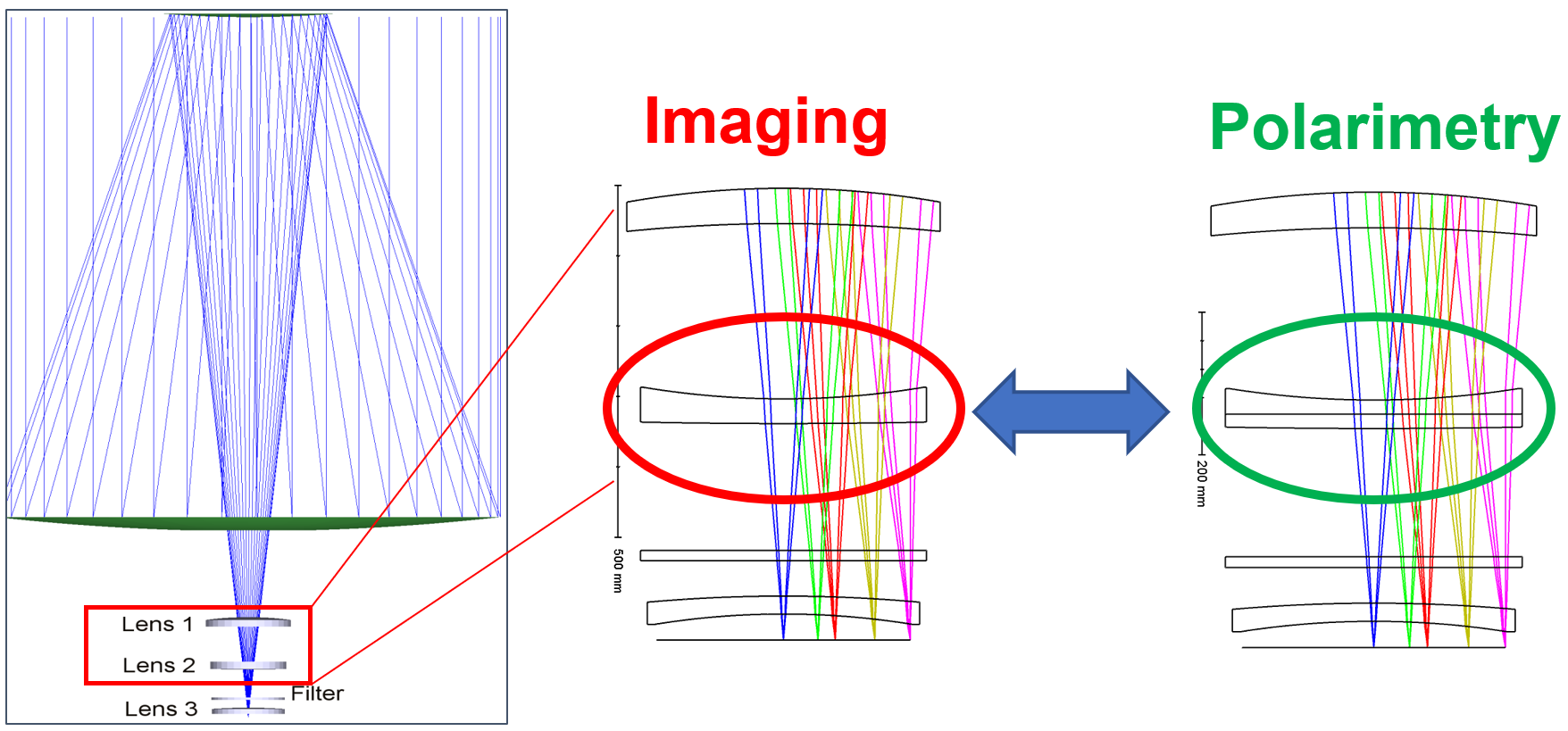}
\end{tabular}
\end{center}
\caption[example1] 
{ \label{fig:NewOptics} 
Left: optical configuration of the telescope, the red rectangle includes the two lenses belonging to the field corrector, which are installed in the old optomechanical device. Center: the unchanged future configuration in imaging mode, where the red oval includes the existing lens which will be made exchangeable. Right: the future configuration in polarimetric mode, where the green oval includes a sandwich of two half-lenses with a polaroid filter in the middle.}
\end{figure}

\section{Science cases for the wide-field}
The high resolution and large FoV of VSTPOL will also allow exploration of many other science cases with a good spatial resolution and an unprecedented coverage of the sky. This is by far the most innovative part of the project. 

In astronomy, polarized emission is mainly produced by specific emission processes, as synchrotron, or often by scattering. In this latter case, the scattering medium can be characterized, deducing, for example, the composition and size of dust grains. This enables us to explore a host of interesting astronomical objects such as comets and asteroids. Exploring their compositions gives insights into their formation, and therefore indirectly provides insights into the formation of our own solar system and galaxy. Examples of scattering based science cases are:

\begin{itemize}
    \item Extending the database of polarimetric asteroids observations
    \item Mapping starlight polarization due to magnetically aligned dust grains in the Milky Way and Magellanic Clouds
    \item Polarimetric surveys of polarized stars, nebulae and galaxies
    \item Monitoring of zodiacal dust and dust clouds in Sun-Earth and Earth-Moon Lagrange points
\end{itemize}

\paragraph{Asteroids and comets}

Polarimetry allows for the classification of asteroids \cite{Masieroetal2012}, as the polarimetric properties of S-type asteroids are remarkably different from the ones of C-type ones. The possibility to estimate the asteroids albedo - and therefore their size - through polarimetric measurements is a remarkable opportunity, as photometry alone cannot distinguish a small and bright object from a large and dark one. A VSTPOL survey of the ecliptic region could increase by a factor of $\sim 10$ the number of V$<20$ asteroids with polarimetric measurements. A definite breakthrough in the field. 

Comet polarization studies is a very promising field. Yet, most current polarimetric observations of comets refer to the region around their nucleus while the nucleus itself is hidden by a coma. Polarimetric characteristics are dominated by the dust ejected by the nucleus and depend on size/size-distribution, shape/structure/porosity, and composition. Models are successful at reproducing single colour polarization curves but are unable to explain their dependence with wavelength. A large FoV offers the possibility to probe the evolution of dust grains and measure the polarization of background stars observed in transmission, substantially enriching the diagnostic content of the polarimetric observations. Finally, VSTPOL allows one to effectively distinguish dust- from gas-rich comets.

\paragraph{Mapping starlight polarization.} Mapping of the starlight polarization of sufficiently large regions of the Milky Way and Magellanic Clouds is a very ambitious research project and requires a large field of view polarimeter with adequate collecting area. The magnetic field is an important component of the Milky Way general interstellar medium (ISM), its energy density being comparable to those of the ISM gas and cosmic ray components. Traditionally, radio (synchrotron emission and Faraday rotation) and now sub-mm (Planck) observations have provided the bulk of the large scale B-field structure, with optical/NIR (starlight) polarization data being still rather limited in comparison. However, with the availability of the GAIA astrometric results (i.e. distances), extensive, wide-field observations of starlight polarization should be able to uniquely probe into the 3-D structure of the field, as well as inform on the dust distribution. A consistent picture of the ISM, in which an optimized dust model and B-field structure can explain the observables over the radio, sub-mm (Planck) and optical/NIR, are then within reach. 

\paragraph{Polarimetric surveys.} In many cases we need to study a large set of objects spread in a relatively large field of view. A typical case is provided by star-forming regions, with the goal, for instance, of constraining and measuring the influence of magnetic fields in star-formation regions. More generally, polarimetric surveys can help in creating a 3D map of the magnetic field, dust grain size, and composition of the interstellar medium. And, finally, the very ambitious plan to measuring the polarization of the Cosmic Background radiation could also be feasible.

\begin{figure} [ht]
\begin{center}
\begin{tabular}{c} 
\includegraphics[height=9cm]{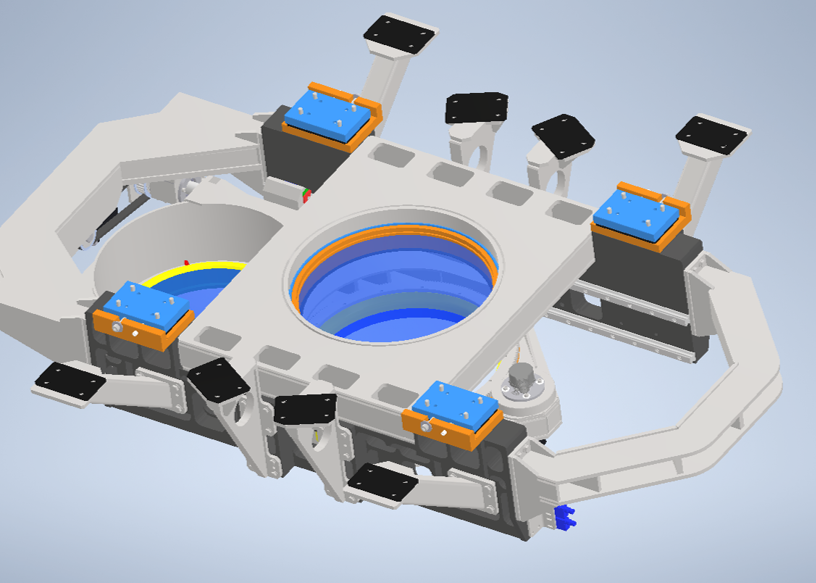}
\end{tabular}
\end{center}
\caption[example1] 
{ \label{fig:NewMech1} 
New mechanical design, top view.}
\end{figure} 

\begin{figure} [ht]
\begin{center}
\begin{tabular}{c} 
\includegraphics[height=9cm]{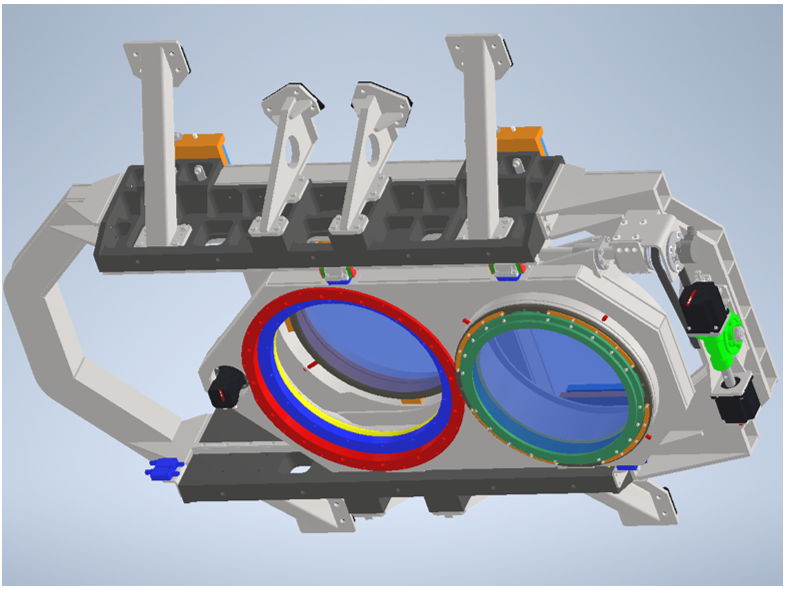}
\end{tabular}
\end{center}
\caption[example1] 
{ \label{fig:NewMech2} 
New mechanical design, bottom view.}
\end{figure}

\section{Design}
\label{sec:design}  
The project comes from an idea proposed by an international consortium including some of the authors of this paper. In 2022 it was adopted by some INAF colleagues who participated to a national call within the Italian National Recovery and Resilience Plan (PNRR) and got funds to realize it, keeping the collaboration with the international team.

The ideal implementation would be a beam-exchange polarimeter consisting of a rotating retarder wave plate in front of a polarizing beam-splitter. Unfortunately, no solution was found for such a beam-splitter that would cover a major part of the field without overlap between the beams. Therefore, the very simple solution with a rotating polarizer was chosen, which not only rejects half of the light and limits our observables to only linear polarization, but also renders the polarimetry sensitive to any temporal variations, like variable seeing and sky transparency. These effects cause a quasi-random offset of the zero point of fractional linear polarization. Fortunately, owing to the large field of view, and a-priori knowledge from the GAIA catalogue, there are always sufficient stars that can be presumed unpolarized in every observation, which can be used as input to a data-driven calibration for these temporal effects. Our simulations using actual OmegaCAM data demonstrate that such corrections can be performed down to fractional polarization levels of $\sim10^{-3}$.

The polarimetric mode for the VST will be implemented through the insertion of a rotating polaroid filter in the optical path, working in the 400-700 nm wavelength range, i.e. with BVgr filters. The polarimetric data will be computed through the combination of several imaging data recorded after rotations of the polaroid filter to given angles, like e.g. [0, 45, 90, 135] or [0, 60, 120].

The optics in front of the polarizer add polarization effects of their own, in the forms of induced polarization, polarization rotation, and cross-talk from linear polarization into unmeasurable circular polarization. The induced polarization is expected to grow with the field angle away from the optical axis, comparable to the well-characterized induced polarization of VLT/FORS\cite{Patat_2006}. We will continuously calibrate these (mostly static) effects at the $\sim10^{-3}$ level using the unpolarized stars in every observation. Due to stresses in the lenses, there may be a small amount of birefringence that modifies the incoming linear polarization as a function of wavelength and position on the detector. We will calibrate for these effects using polarized standard stars.

The current imaging mode of the VST shall be kept with no modifications, i.e.~the existing field-corrector shall be usable in normal imaging mode as in the current conditions. 

More details on the management of the project and the systems engineering aspects are given in [\citenum{Savarese24_1}].

\subsection{The starting situation}
The VST optical design is shown in Fig.~\ref{fig:Telescope}. The VST is a 2-mirror telescope equipped with a field corrector composed of 3 lenses. The third lens is physically hosted in the OmegaCAM instrument, as well as the filters. The first 2 lenses of the corrector (hereinafter: 2-lens corrector) are physically in the telescope. They are enclosed within the red rectangle in Fig.~\ref{fig:Telescope}. They do not follow the field rotation, as of course the instrument does.
The 2-lens corrector mechanical support is connected to the back side of the primary mirror cell, as shown in Fig.~\ref{fig:CurrentMechanics} on the left. The current mechanical support hosts the 2-lens corrector and a second group of lenses (an Atmospheric Dispersion Corrector, currently unused), which can replace the 2-lens corrector in the optical path through a linear exchange mechanism. There is no accessibility to the whole device, as shown in Fig.~\ref{fig:CurrentMechanics} on the right. It is in a sandwich between the primary mirror cell on the top and the VST adapter-rotator system, not mounted in the left side of the picture, on the bottom. The OmegaCAM instrument is connected to the rotator, obstructing any access. The access to the 2-lens corrector is only possible after removal of:

\begin{itemize}
    \item the OmegaCAM instrument 
    \item the VST adapter-rotator
\end{itemize}

This poor accessibility has been taken into account in the design of the new system, by implementing redundancies to ensure the telescope can keep working also in case of failures.

\subsection{The new implementation}
The VSTPOL setup will allow the VST telescope to work in two different observing modes:

\begin{itemize}
    \item Normal Imaging (photometry)
    \item Linear Polarimetry
\end{itemize}

The fundamental guideline is that the telescope must preserve all its current features and add the polarimetric capability. This will be achieved thanks to the capability of changing the optics in the optical path.

VSTPOL will be implemented by replacing the electro-opto-mechanical system shown in Fig.~\ref{fig:CurrentMechanics} on the left. The complete replacement will minimize the telescope downtime due to the intervention. In fact, the alternative option of a reworking on site of the existing parts would be extremely difficult, and the period of stop would be unpredictably large.

The lenses of the existing 2-lens corrector shall be removed from the current mechanics and installed in the new one. The new system shall allow the exchange between the existing 2-lens corrector and the new polarimetric optics. This will be implemented by:

\begin{itemize}
    \item leaving the first lens (L1) unchanged, in a fixed position
    \item building an exchange mechanism to insert alternatively in the optical path either the old second lens (L2) to work in photometric mode as usual, or a new L2, actually a sandwich composed of 2 halves lenses with a polaroid filter inside
\end{itemize}

This design comes from a study where several options have been considered in terms of reuse of the old optics and position of the polaroid filter. The final decision is the outcome of a trade-off between performance, feasibility and cost. 

\begin{figure} [ht]
\begin{center}
\begin{tabular}{c} 
\includegraphics[height=10.5cm]{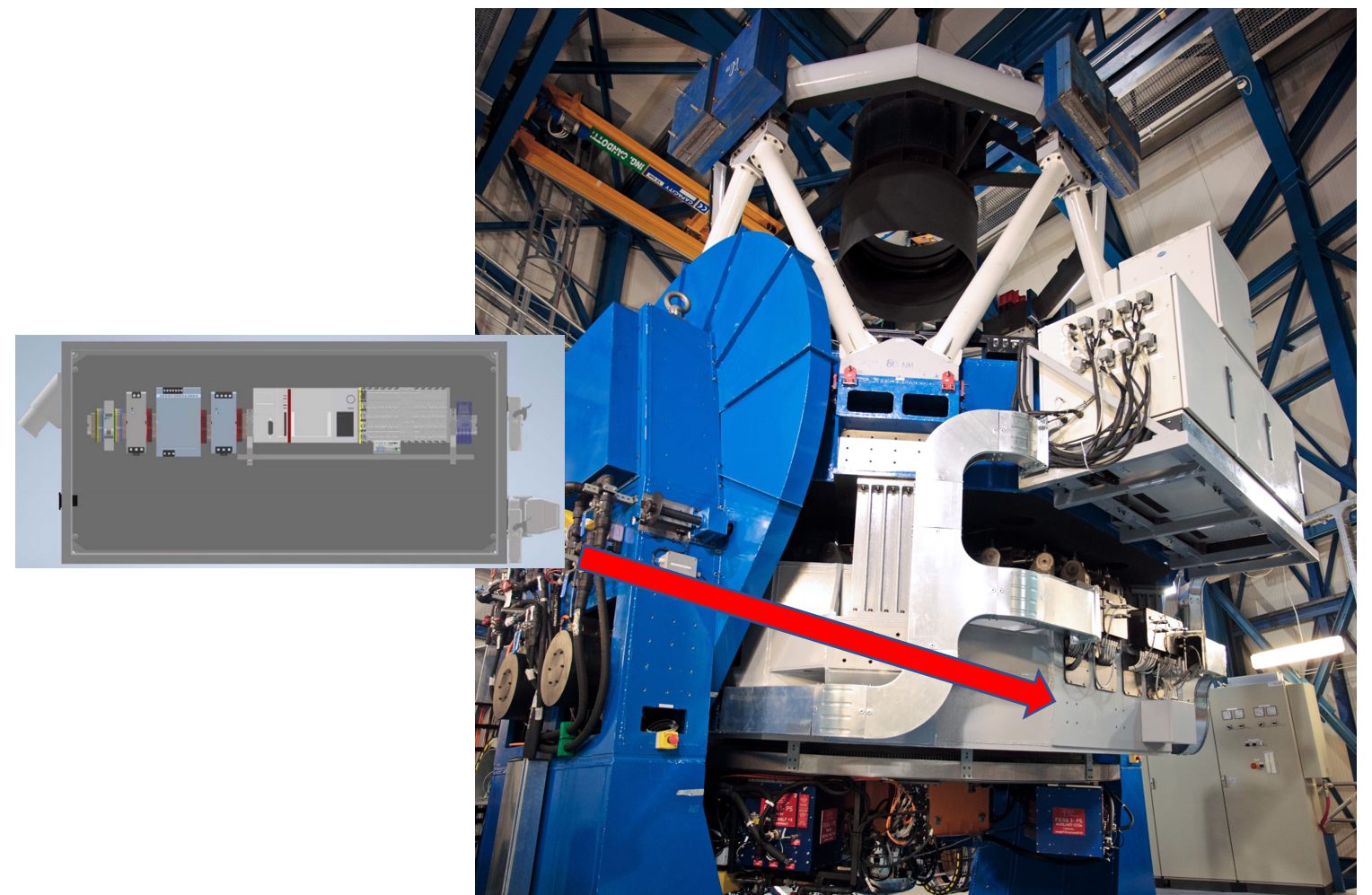}
\end{tabular}
\end{center}
\caption[example1] 
{ \label{fig:NewElect} 
Design of the control electronics box and its future location on top of the telescope.}
\end{figure} 

\begin{figure} [ht]
\begin{center}
\begin{tabular}{c} 
\includegraphics[height=9cm]{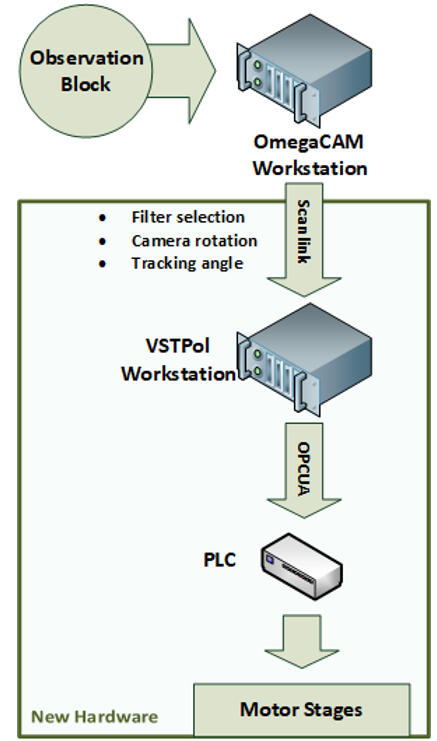}
\end{tabular}
\end{center}
\caption[example1] 
{ \label{fig:NewCtrl} 
Sketch of the upgrade of the control architecture.}
\end{figure}

\section{Optical Design}

In photometric mode, the L1 is slightly axially moved with respect to the current optical VST configuration. This helps to compensate aberrations in both observing modes.  A slight secondary mirror refocus ($\sim$50$\mu$m) is needed to compensate for the lens position. 

In polarimetric mode, the field corrector is modified with a new setup including the polarimetric filter. The large polarizer is 1.6 mm thick with a broadband antireflective coating, characterized by 44\% transmission and 99.98\% efficiency in the 400-700 nm wavelength range. The polarimetric filter is installed within the second element, splitting the second lens of the two-lens design. 
The first lens L1 is reused.  Performance are as good as in the current 2-lens configuration.
A drawback of this option might have been the thickness of the two lenses. The layout has been conceived with a thicker L2, while maintaining L1 unchanged. This proved to be feasible, within a certain extent.
Each of the halves of the new L2 combined lens has a center thickness of at least 20 mm. This value has been considered safe enough for the manufacturing, testing, and handling of the two lenses. 
A slight secondary mirror refocus w.r.t. the photometric mode will be needed, to compensate for the different glass thickness of L2 in the two different configurations.

Summarizing, this new design reuses the L1 lens, that will be mounted at a slightly different axial position with respect to its current one, 2.5 mm closer towards the VST focal plane, in a fixed holder. The current L2 lens and the new polarimetric L2 combined lens will switch between the two optical configurations. Both L2 lenses will have the same exit vertex axial position of the current design, to avoid any collision with other optomechanical subsystems (e.g. telescope probe arm).

\section{Mechanical Design}
The mechanics shall support the L1 fixed lens, always positioned in the optical path (Fig.~\ref{fig:NewMech1}), and the two interchangeable optical elements, the original L2 lens and the “L2 combined polarimetric lens”, installed on the exchange mechanism (Fig.~\ref{fig:NewMech2}). This linear mechanism will select either the original L2 lens, setting the telescope for the photometric mode, or the polarimetric lens setting the telescope for polarimetry.

As the new mechanical support will replace the old one, the mechanical interfaces towards the mirror cell will be the same.

The L2 polarimetric lens needs to be rotated to different angles to get a set of images with different polarization properties, allowing to disentangle the different polarization components. Also, it has to follow the movement of the OmegaCAM instrument that is installed on the instrument rotator to compensate for the field rotation. These two requirements shall be accomplished by a motorized rotation device for the polarimetric lens. Of course the rotation is only needed for the polarizer optics; the original L2 lens does not need to rotate and will be fixed to the structure of the exchange mechanism.

\section{Control Electronics}
The control electronics\cite{Colapietro24} is designed following the new ESO standards. A Beckhoff PLC architecture based on Commercial Off-The Shelf (COTS) components is adopted. Two functions shall  be controlled:

\begin{itemize}
    \item The linear exchanger mechanism 

It enables the switch between the traditional imaging and the new polarimetric mode. The mechanism will be driven by a stepper motor connected, through a  transmission system, to a screw implementing the linear motion. 
A rotary absolute encoder connected on axis with the screw will provide feedback on the actual position of the moving element. 
The device will be equipped with two couples of end-of-travel switches, implementing a redundancy. 
A spare stepper motor will be installed as well, to allow for the repositioning of the original 2-lens field corrector in the optical path in case of failure of the nominal motor. These redundancies are needed because of the poor accessibility of the device: in case of troubles the only way to access it is the removal of the OmegaCAM instrument and the telescope adapter-rotator, a long and unpleasant work.

    \item The polaroid rotation mechanism 
    
It is needed to move the polaroid filter to different angles, getting images with different polarization properties. Also, it must enable tracking of the polaroid, following the movement of the instrument that compensates for the field rotation.
It is driven by a stepper motor through a gear system. A rotary absolute encoder is connected to the system through a gear to provide position feedback. 

All the control electronics is hosted in a wall-mountable box to be installed on the telescope (Fig.~\ref{fig:NewElect}): the enclosure shall be equipped with a cooling circuit, actively controlled by the PLC.
\end{itemize}

\section{Control Software}
The control software must be integrated into the existing OmegaCAM instrument software. Indeed, the VSTPOL optomechanics will be a part of the telescope, fully decoupled by the instrument. Nevertheless, logically the handling of the polarimetric mode is part of the instrument software, because the instrument remains the same, but in polarimetry simply works with the addition of a rotating polaroid filter. Therefore the control software shall allow the astronomer to work either in pure imaging, as usual, or in polarimetric mode.

Hence, the OmegaCAM instrument software\cite{Baruffolo02,Baruffolo04} must be upgraded. This has some challenges because the original software was based on different hardware platforms (VME), discontinued because no longer supported. The new PLC-based hardware platform is handled very differently at software level, so the upgrade will unavoidably mix old and new hardware and software solutions. The control architecture is sketched in Fig.~\ref{fig:NewCtrl} and more extensively described in [\citenum{Savarese24_2}].

\section{Conclusions}
VSTPOL is a modification to the telescope to add unique polarimetric capabilities to the VST, rather than a project for a new instrument. The new design will preserve the current capabilities of telescope and camera, adding new features for the exploration of new science.

According to the current schedule, the device will be ready for acceptance in Italy within end of 2025, the installation and commissioning in Chile will follow in 2026.

\acknowledgments 
The authors acknowledge the support from the Next Generation EU funds within the National Recovery and Resilience Plan (PNRR), Mission 4 - Education and Research, Component 2 - From Research to Business (M4C2), Investment Line 3.1 - Strengthening and creation of Research Infrastructures, Project IR0000012 – “CTA+ - Cherenkov Telescope Array Plus”.

\bibliography{VSTPOL} 
\bibliographystyle{spiebib} 

\end{document}